\documentclass[aps,reprint,prb,twocolumn,superscriptaddress]{revtex4-1}
\usepackage{color}
\usepackage{graphicx}

\begin{document}

\title{Quasi-bound States and Evidence for a Spin 1 Kondo Effect in Asymmetric Quantum Point Contacts}
\author{Hao Zhang}
\affiliation{Department of Physics, Duke University, Physics Building, Science Drive, Durham, North Carolina 27708, USA}
\author{Phillip M. Wu}
\affiliation{Department of Physics and Applied Physics, Stanford University, Stanford, California 94305}
\author{Albert M. Chang}
\affiliation{Department of Physics, Duke University, Physics Building, Science Drive, Durham, North Carolina 27708, USA}

\date{\today}

\begin{abstract}

Linear conductance below $2e^2/h$ shows resonance peaks in highly asymmetric quantum point contacts (QPCs).  As the channel length increases, the number of peaks also increases.  At the same time, differential conductance exhibits zero bias anomalies (ZBAs) in correspondence with every other peak in the linear conductance.  This even odd effect, observable in the longer channels, is consistent with the formation of correlation-induced quasi-localized states within the QPC. In rare cases, triple peaks are observed, indicating the formation of a spin one Kondo effect when the electron filling number is even. Changing the gate voltage tunes this spin triplet to a singlet which exhibits no ZBA. The triple-peak provides the first evidence suggestive of a spin singlet triplet transition in a QPC, and the presence of a ferromagnetic spin interaction between electrons.

\end{abstract}

\maketitle

The conductance of a quantum point contact (QPC), a quasi-one-dimensional quantum wire, is quantized in units of $G_0 = 2 e^2/h$. \cite{vanWees,Wharam} This quantization can be understood within a single-particle picture: the density of states cancels the velocity (up to a multiplicative constant) yielding a unit conductance $G_0$ for each subband (accounting for both spin projections). Thus, the total conductance only depends on the number of occupied subbands. However, an additional bump near $0.7 G_0$ has drawn much attention as an indication of nontrivial interaction physics.\cite{Thomas} Among the many theoretical proposals \cite{Micolich}, a possible explanation is the formation of a quasi-bound state due either to correlation effects, \cite{Meir, Rejec, Guclu} or to a momentum mismatch.\cite{Song} This 0.7 feature remained the only clear-cut manifestation of possible interaction effect until recent years, in spite of the often raised possibility of spin-polarization, which has never been convincingly established in n-type devices. Additional intriguing behaviors are now being uncovered. First, a modulation/destruction of the lowest conductance plateau was reported.\cite{Hew}  Second, within the past year, Wu {\it et al.} reported evidence for quasi-bound states in QPCs with a highly asymmetric geometry, manifested in the presence of sharp resonances and modulation/suppression of the lowest plateau, as well as non-Fermi-liquid temperature evolution of resonance peaks below $G_0$.\cite{Wu}  A natural question arises regarding: Are the observed features caused by intrinsic electron-correlation, brought about by the unusual device geometry, or do they arise from disorder caused by impurities and lithographic imperfections? To address this important issue, we report a systematic study on the dependence of these features, as well as non-linear transport properties, on channel length.

We will first demonstrate that the number of conductance resonances increases with an increase in the channel length.  This behavior has long been predicted
based on quasi-bound-states formation,\cite{Kirczenow,Marel,Szafer} but yet to be observed experimentally. Moreover, the quantized conductance plateaus can be modulated/suppressed by tuning gate voltages. We ascribe these two features to the formation of the quasi-bound states due to the electron correlation effect.\cite{Song, Guclu, Berggren, Akis, Abhijit} In addition to the linear conductance, the differential conductance (dI/dV) was systematically studied. The zero bias anomaly (ZBA) in the dI/dV can be correlated to the linear conductance peaks. In most cases for longer channels $\ge 500~nm$, the ZBA is observed for every other linear conductance peak, reminiscent of the even-odd Kondo effect in quantum dots,\cite{GG, Jeong} Thus, each successive linear conductance peak corresponds to adding an electron into the quasi-bound states.  However, when the filling number is even, some dI/dV curves exhibit triple peaks near zero bias. This suggests the formation of a spin 1 (triplet) Kondo state instead of a spin 0 (singlet) state. This spin 1 Kondo may arise due to the presence of a ferromagnetic electron spin coupling.\cite{Song}  By tuning the gate voltage, it was possible to remove the triple-peak and cause a transition to a singlet state.  

\begin{figure}[t]
		\includegraphics[width=8.8cm, height=6.2cm]{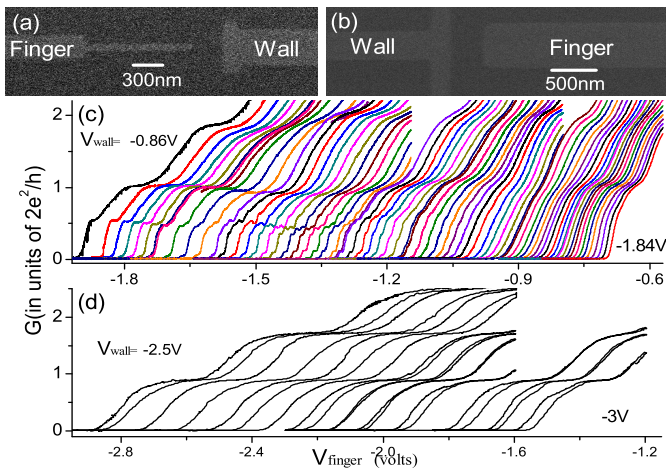}
		\caption{(Color online)(a),(b) SEM images of two asymmetric QPCs which have $100~nm$ and $500~nm$ channel length, respectively. (c) Conductance of an asymmetric QPC with  $500~nm$ channel length and $300~nm$ gap. $V_{wall}$ is $-0.86~V$ for the leftmost curve(black) and $-1.84~V$ for the rightmost (red), and (d) conductance for a symmetric QPC with $500~nm$ channel length and $450~nm$ gap. $V_{wall}=-2.5~V$ for the leftmost and $-3~V$ for the rightmost. $T=4.2~K$.}
		\label{geometry}
\end{figure}

Both symmetric and asymmetric QPCs, with split-gate gaps of $450~nm$, $300~nm$ and $250~nm$, were fabricated by electron beam lithography, evaporation of the Cr/Au surface gates, and lift-off. The nominal channel lengths, defined by the gate length, varied from $100~nm$ to $\sim 1000~nm$.   In the GaAs/AlGaAs heterostructure crystal, the two dimensional electron gas (2DEG) is located at a shallow depth $80~nm$ below the surface. The carrier density and mobility are $3.8\times 10^{11} cm^{-2}$ and $9\times 10^5 cm^2/Vs$, respectively, giving a mean-free-path $\sim 9~\mu m$. An excitation voltage $V_{ac}=10~\mu V$ at $17.3~Hz$ was applied across the QPC, and the current was measured in a PAR124A lock-in amplifier after conversion to a voltage using an Ithaco 1201 current preamplifier.

Fig.1(a)(b) show typical gate geometries for the asymmetric QPCs. Unlike our symmetric QPCs which have two symmetric finger split-gates, we replace one finger with a relatively long wall to geometrically introduce asymmetry. The width of the finger gate defines the lithographic channel length, and the gap is the separation between the wall and finger.  The QPC is formed by applying negative voltages to these gates to deplete the 2DEG underneath. Fig. 1(c) shows linear conductance traces  for an asymmetric QPC, while Fig. 1(d) for a symmetric QPC at 4.2K. For every asymmetric QPC, each conductance trace was measured by fixing the wall voltage ($V_{wall}$), while sweeping the finger voltage ($V_{finger}$). For every symmetric QPC, conductance was measured by fixing one finger voltage (also labeled as $V_{wall}$) while sweeping the other. Different traces in Fig. 1(c) or (d),  correspond to different fixed $V_{wall}$.  The symmetric QPC (Fig. 1(d)) shows well-defined quantized plateaus throughout all gate voltages indicating ballistic transport. For the asymmetric QPC(Fig. 1(c)), besides the conductance quantization which exhibits modulation/suppresion, conductance resonances are observed below the first quantized plateau. This conductance resonances shift positions when $V_{wall}$ is tuned from one trace to another. The plateau value also oscillates about $G_0$ as $V_{wall}$ is tuned. Moreover, the first plateau is suppressed near $V_{finger}\sim -0.9~V$.  Our previous work\cite{Wu} reported this suppression of $G_0$ plateau for $100~nm$ channel length asymmetric QPCs down to $300~mK$. Here we find that it is also observable in longer channels. Intriguingly, below $G_0$, a strong resonance series near $0.5G_0$ is present at this high temperature. However, this $0.5G_0$ series was only present in two of our asymmetric QPCs, indicating a sensitivity to the precise channel shape.

Dopant impurities or lithography imperfections on the gates can cause backscattering, leading to conductance resonances\cite{McEuen,Sfigakis}. To rule out these possibilities, the symmetric QPCs are utilized as a control group. More than 30 asymmetric QPCs and 15 symmetric QPCs, with different channel lengths and gaps, are measured at $4.2~K$ or $3~K$. They all qualitatively agree with the data in Fig.1(c)(d): all asymmetric QPCs tend to show conductance resonances and modulation/suppression of plateaus while all the symmetric QPCs show no (or much fewer) resonances and a reduced (or a lack of) modulation of plateaus. It is highly unlikely that only the asymmetric QPCs have impurities or lithography imperfections while the symmetric QPCs do not. Thus, the two main features: (most) conductance resonances and the modulation/suppression of plateaus are intrinsic.

\begin{figure*}[ht]
		\centering
		\includegraphics[width=17.5cm, height=6.5cm]{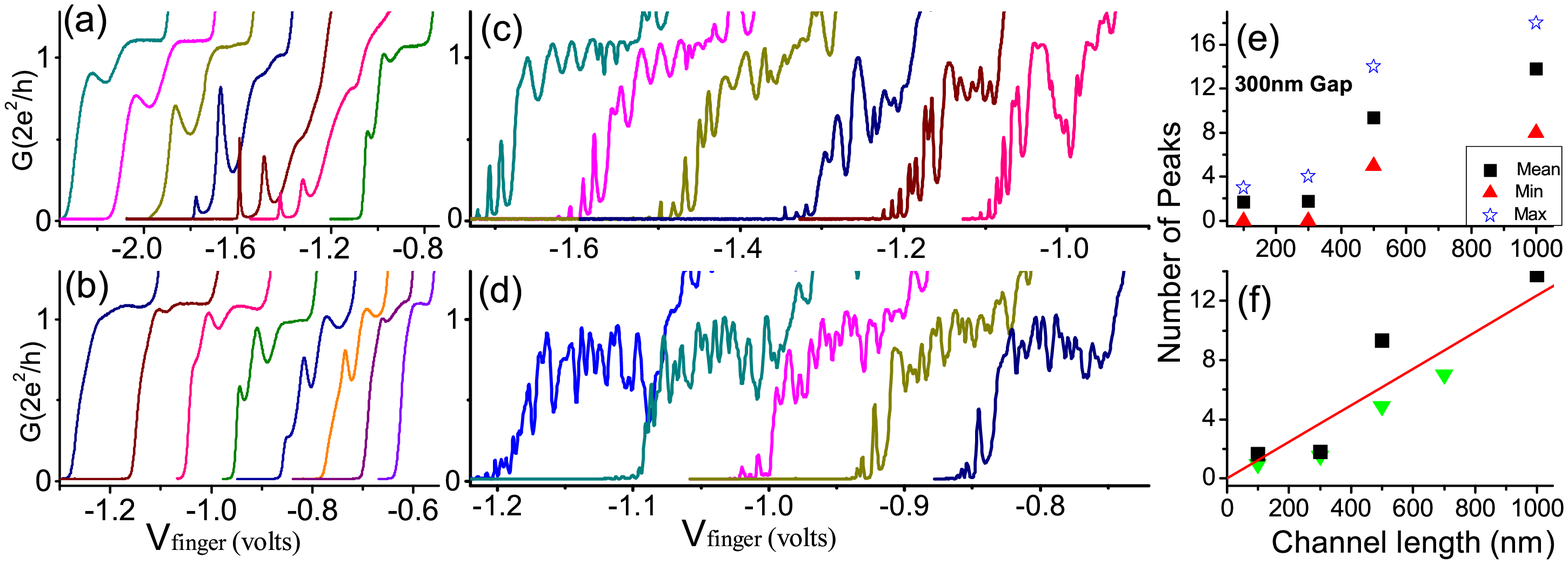}
		\caption{(Color online) Linear conductance of four typical asymmetric QPCs at $300~mK$. (a)-(d) conductance of 100, 300, 500, and 1000 $nm$ channel length asymmetric QPCs with $300~nm$ gap.  Correspondingly, the leftmost curve has $V_{wall}= -0.91, -0.71,-0.59$ and $-0.59~V$ while the rightmost $V_{wall}= -1.47, -1.83, -1.56$ and $-1.23~V$.  Number of peaks below $G_0$ vs. channel length: (e) for $300~nm$ gap, and (f) mean number for $250~nm$ and $300~nm$ gap.}
	\label{lowtempcond}
\end{figure*}

At the lower temperature of $\sim 300~mK$ shown in Fig. 2, the curves for asymmetric QPCs develop many oscillations. For clarity, in Figs. 2(a)-(d) we present $\sim 1/8$ of the curves measured for the $300~nm$ gap QPC set. The $250~nm$ gap QPCs exhibit similar resonances. Note that as the length increases, the number of peaks below the first plateau tends to increase, while at the same time, the typical peak width narrows.  Moreover, the spacing between peaks close to threshold is consistent with the Coulomb charging energy estimates from the channel capacitance to the surroundings.  Overall, thermal cycling laterally shifts the curves and changes the details, but the main behavior remains consistent, further indicating intrinsic behavior. The suppression of the $1 \times G_0$ first plateau is clearly seen in (a)(c), but weaker in (b).  For each channel length we tabulated the number of peaks below $G_0$ for each of the $\sim 50$ curves, and present the statistics in Fig. 2(e)(f).  In (e) each (black) square point represents the mean number of peaks for a given channel length in the $300~nm$ gap set. (Blue) stars and (red) triangles represent the maximum and minimum number, respectively.  In (f), the mean numbers for both $300~nm$ and $250~nm$ gap are plotted. The resonances versus channel length, reflected both in the raw traces and the statistics in Fig. 2, clearly demonstrate that the number of resonances increases with, and is approximately proportional to the length. 

Besides the thermal cycling and the contrasting behavior with symmetric QPCs, we furhter stress the intrinsic nature of the conductance resonances and modulation/suppression of plateaus based on the following. The presence of impurities can be a cause of resonances and lead to a degradation of the quantized plateaus.\cite{Baranger} In our case, degradation of plateau quantization is not observed. At $300~mK$, the plateaus are clearly present, albeit with resonance oscillations superimposed.  As for the lithography imperfection, which may randomly leave bumps on the gate, probabilistically it is highly unlikely that they are present in all the short channel ($100$ or $300~nm$) devices; bumps have typical dimensions of $20-50~nm$ length-wise, and occur every $\sim 300~nm$ (based on SEM imaging).

To gain insight, we may motivate presence of the conductance peaks within a single-particle picture, as resulting from quasi-bound states.  When scanning $V_{finger}$, the  relative Fermi level, defined as the distance from the first subband level to the chemical potential, changes, along with the Fermi wavelength $\lambda_F$.  In the simplest scenario, when $N\cdot\frac{\lambda_F}{2}=L$, with N is a positive integer and L the channel length, a tunneling resonance (quasi-bound state) will form.  Longer channels have quasi-bound states more dense in energy, so more conductance resonances are observed. This single particle picture is based on a momentum mismatch at the channel entrance and exit. In a traditional symmetric QPC typically fabricated in 2DEGs residing much deeper than 80 nm below the surface, the resutling adiabatic smooth potential profile reduces momentum mismatch, minimizing the back-scattering necessary for the formation of the quasi-bound states. \cite{Szafer} Recent theoretical calculations\cite{Guclu, Abhijit} have shown that without considering electron correlations, even if the gates induced bare potential were sharp at the QPC entrance, the effective potential profile can readily be smoothed out due to screening. Thus the conductance resonances we observe can be explained only if correlation is included. Exact diagnolization\cite{Song}, spin density functional theory (SDFT)\cite{Berggren} and quantum Monte Carlo(QMC) calculations\cite{Guclu} indicate that when correlation is included, quasi-bound states can form. SDFT and QMC show that electron correlation can induce potential barriers, causing the backscattering needed for forming quasi-bound states. Intriguingly, using SDFT, Akis and Ferry found that these correlation induced barriers can cause conductance resonances and modulation/suppression of quantized plateaus,\cite{Akis} in a QPC/2DEG 70 nm below the surface. Their calculations produced conductance traces qualitatively similar to the data shown here. As may be expected, SDFT\cite{Berggren} and QMC\cite{Guclu, Abhijit} demonstrate that it is easier to form correlation-induced barriers when the gate potential profile is sharper. This suggests that even for symmetric QPCs, electron correlation may induce barriers and conductance resonances, as long as the gate potential profile is sharp enough. Resonances are observable in some of our symmetric QPCs, although the number is much fewer. Their presence may be due to the fact that our 2DEG is buried at a relatively shallow depth ($80nm$) compared to other groups', causing a sharper potential profile.  In an asymmetric QPC defined by surface oxidation, which should also yield a sharper potential profile, similar reproducible resonance features was observed by Senz {\it et al.}.\cite{Senz}

\begin{figure*}[t]
		\centering
		\includegraphics[width=17.5cm, height=6.5cm]{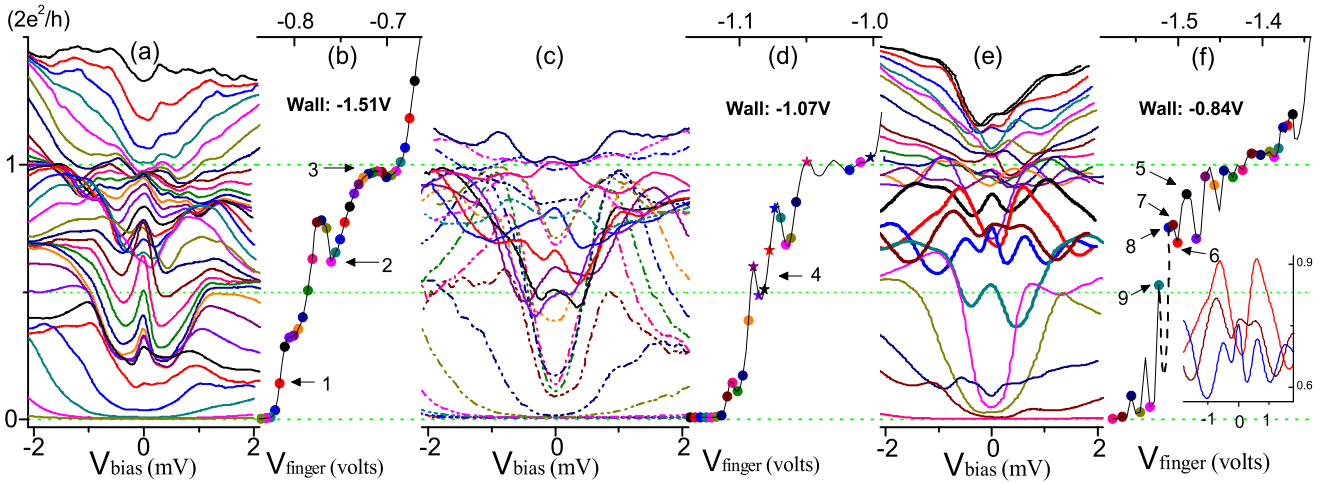}
		\caption{(Color online) Differential and linear conductance of three asymmetric QPCs at $300~mK$. (a) dI/dV for $300~nm$ channel length, $300~nm$ gap with a fixed $V_{wall}=-1.51~V$, while sweeping $V_{bias}$ (x-axis) and $V_{finger}$ (different colors). (b) is a plot of the center (linear conductance) of (a) as a function of $V_{finger}$. (c),(d) dI/dV for $500~nm$ channel, $250~nm$ gap. Star points exhibit ZBA and circular points do not. (e),(f) dI/dV for $500~nm$ channel, $300~nm$ gap. Conductance values $G_0$ and $\frac{1}{2}G_0$ are indicated by horizontal dotted lines. In (f), the dashed line connecting points 8 and 9 is from the linear conductance data in figure 2(c) and accounts for a missing point in the center cut of 3(e). Inset highlights the singlet-triplet transition curves in (e). }
	\label{DIDV}
\end{figure*}
 
In Fig. 3, we present the nonlinear conductance (dI/dV) as a function of source-drain bias $V_{bias}$.  For clarity, only 50\% of the traces is included for each QPC.  Panel (a) shows typical behavior in a short channel asymmetric QPC. $V_{wall}$ is held fixed while $V_{finger}$ is increased in steps (bottom to top). Each curves for a different $V_{finger}$, is shown with a different color online. In panel (b), the center cut of (a), equivalent to the zero-bias linear conductance, represented with a symbol of the same color, is plotted versus $V_{finger}$. The black line connecting all symbols, including those not shown, yields the zero-bias linear conductance versus $V_{finger}$ at the fixed $V_{wall}$. This curve is consistent with the corresponding linear conductance curve measured separately in Fig. 2.

Zero bias peaks (ZBPs) in QPCs is a nonlinear conductance peak near zero bias discovered by Cronenwett et al.\cite{Cronenwett}, associated with the zero bias anomaly (ZBA).  In their work, the temperature and magnetic field dependence of the ZBA, analogous to the Kondo effect in quantum dots (QDs),\cite{GG, Jeong} suggest the possibility of a quasi-localized state (a localized magnetic impurity). In Fig. 3(a), the ZBA begins at the (red) curve (arrow 1 in (b)), and continues up to the first plateau (arrow 3 in (b)), similar to other groups' results.\cite{Cronenwett,Sfigakis,Folk} The amplitude of ZBA has a minimum at the second valley in (b) (arrow 2). Three other short channel QPCs exhibit similar features.

Figs. 3(c)(d) show the near even-odd effect which is, instead, observed in our long channel asymmetric QPCs. In a QD, ZBP is observed for odd electron filling number (non-zero magnetic moment), while in most cases, no ZBP is observed for an even filling.\cite{GG} This even-odd effect in QDs gives rise to the ZBP in every other Coulomb blockade valley.  In Fig. 3(c), the solid curves exhibit the ZBA whereas the dotted curves do not. This ZBA and non-ZBA alternation is differentiated in (d) using different symbols (stars for ZBA and circles for no ZBA). QPCs are normally thought of as open systems, while the even-odd effect occurs in confined systems, e.g. QDs. However, based on the previous discussion, electron correlation-induced barriers lead to quasi-bound (quasi localized) states, causing the even-odd effect. Our ZBA does not strictly follow the even-odd law, however. Occasionally it can appeared to be even-odd-odd or even-even-odd. Near threshold, such behavior can result from the small electron tunneling rate to the leads ($\Gamma$), which suppresses the Kondo temperature exponentially, rendering the ZBA unobservable.  On the other hand, near the $1 \times G_0$ plateau, the channel is quite open (nearly an open system), and it becomes meaningless to define filling number.  This near even-odd effect is, observable in the longer channels ($500~nm$, $700~nm$ and $1000~nm$).

One feature in Fig. 3(e) is particularly intriguing.  Two dI/dV curves (brown and blue online) of intermediate conductance show triple peaks near zero bias.  Following the linear conductance in (f) and cross referencing with (e), The (black colored) curve or dot (labeled by arrow 5) shows a single ZBP, suggesting an odd filling number and a spin 1/2 state. Going past the peak in the linear conductance, the curve (red, arrow 6) shows no ZBA but two big bumps on the sides, suggesting that one electron has popped out and now the filling number is even. The next curves (brown, arrow 7 and blue, 8) have triple peaks which is unusual as the filling number is still even.  Going further, the curve (green, arrow 9), has a ZBA suggests the filling number is now odd again.  (Note that the rapidly rising conductance background can shift the peak and valley positions).  The (black and green) curves (arrows 5 and 9) have odd filling number with ordinary ZBAs. In between where the filling is even, curves 7, 8 (also shown in the inset of (f)) with a triple-peak ZBA transition into those with no-ZBA but with a double-side-peak (e.g. 6). We ascribe this to a spin triplet-to-singlet transition based on the following reasoning. For even filling, the ground state can either be a spin-singlet or a triplet, depending on the competition between the exchange interaction ($E_{ex}$) and the orbital level spacing ($\Delta E_{l}$). For the singlet, two electrons occupy the same orbital level, the energy is mainly $E_{ex}$, while for the triplet, two electrons occupy two different orbital states, and the energy is mainly $\Delta E_{l}$. If by tuning $V_{finger}$, $\Delta E_{l}$ happens by chance to exceed $E_{ex}$, the ground state is a singlet with no Kondo ZBA, as shown in arrow 6 (red curve).  However, if $\Delta E_{l}< E_{ex}$, a triplet will be favored (arrow 7, 8 , brown and blue curves). The partial Kondo screening of this spin 1 causes the central ZBP, while the the two side peaks correspond to a triplet-to-singlet excitation occurring via a second-order process. Singlet-triplet transition has been observed in semiconductor QDs\cite{Kogan}, and in $C_{60}$ molecular QDs\cite{Roch}, but not as yet in QPCs. The central ZBP in the triplet regime is usually narrower than the two side peaks\cite{Roch, Kogan, von} while our central peak is only slightly ($\sim 25 \%$) narrower. This is due to the thermal broadening, with $3.5k_B T (\sim 0.1meV)$ roughly the width of the central ZBP. This width is comparable to that in the triplet ZBP obtained by other groups' at a similar temperature\cite{Roch, Kogan, von}.  The rich spin behaviors observed have been suggested in numerical diagonalization calculations.\cite{Song} 

We thank M. Pepper, S. Florens, D. Liu, A. C. Mehta, and H. U. Baranger for useful discussions. Work supported by NSF DMR-0701948.


\begin{references}

\bibitem{vanWees}
B.J. van Wees, H. van Houten, C. W. J. Beenakker, J. G. Williamson, 
L. P. Kouwenhoven, D. van der Marel, and C.T. Foxon, 
Phys. Rev. Lett. {\bf 60}, 848 (1988).

\bibitem{Wharam}
D. A. Wharam, T. J. Thornton, R. Newbury, M. Pepper, H. Ahmed, J. E. F. Frost, 
D. G. Hasko, D. C.  Peacock, D. A. Richie, and G. A. C. Jones, 
J. Phys. C {\bf 21}, L209 (1988).

\bibitem{Thomas}
K. J. Thomas, J. T. Nicholls, M. Y. Simmons, M. Pepper, D. R. Mace, 
and D. A. Ritchie, Phys. Rev. Lett. {\bf 77}, 135 (1996).

\bibitem{Micolich}
A. P. Micolich, Journal of Physics: Condensed Matter {\bf 23}, 
443201 (2011).

\bibitem{Meir}
Y. Meir, K. Hirose, and N. S. Wingreen,
Phys. Rev. Lett. {\bf 89}, 196802 (2002). 

\bibitem{Rejec}
T. Rejec and Y. Meir, Nature {\bf 442}, 900 (2006).

\bibitem{Guclu}
A.D. G\"{u}\c{c}l\"{u}, C.J. Umrigar, Hong Jiang, and Harold U. Baranger, 
Phys. Rev. B {\bf 80}, 201302 (R) (2009).

\bibitem{Song}
Taegeun Song and Kang-Hun Ahn, Phys. Rev. Lett. {\bf 106}, 057203 (2011).

\bibitem{Hew}
W. K. Hew, K. J. Thomas, M. Pepper, I. Farrer, D. Anderson, G. A. C. Jones, D. A. Ritchie, Phys. Rev. Lett. {\bf 101}, 036801 (2008)

\bibitem{Wu}
P. M. Wu, Peng Li, Hao Zhang, and A. M. Chang, Phys. Rev. B {\bf 85}, 
085305 (2012).

\bibitem{Kirczenow}
G. Kirczenow, Phys. Rev. B {\bf 39}, 10452 (1989).
 
\bibitem{Marel}
D. van der Marel and E. G. Haanappel, Phys. Rev. B {\bf 39}, 7811 (1989).
 
\bibitem{Szafer}
Aaron Szafer and A. D. Stone, Phys. Rev. Lett. {\bf 62}, 300 (1989).

\bibitem{Berggren}
I. I. Yakimenko, V. S. Tsykunov and K-F Berggren, J. Phy.,: Condens. Matter {\bf 25}, 072201 (2013).

\bibitem{Akis}
Richard Akis and David Ferry, NSTI-Nanotech {\bf 3}, 240 (2005), ISBN 0-9767985-2-2. 

\bibitem{Abhijit}
Abhijit C. Mehta, PhD Thesis, Duke University (2013). 

\bibitem{GG}
D. Goldhaber-Gordon, H. Shtrikman, D. Mahalu, D. Abusch-Magder,
U. Meirav, and M. A. Kastner, Nature {\bf 391}, 156 (1998).
 
\bibitem{Jeong}
H. Jeong, A. M. Chang, and M. R. Melloch, Science {\bf 293}, 2221 (2001).
 
\bibitem{McEuen}
P. L. McEuen, B. W. Alphenaar, R.G. Wheeler, R. N. Sacks, Surf. Sci. {\bf 229}, 312(1990).

\bibitem{Sfigakis}
F. Sfigakis, C. J. B. Ford, M. Pepper, M. Kataoka, D. A. Ritchie, 
and M. Y. Simmons, Phys. Rev. Lett. {\bf 100}, 026807 (2008). 

\bibitem{Baranger}
John A. Nixon, John H. Davies, and H.U. Baranger, Phys. Rev. B {\bf 43}, 
12638 (1991).

\bibitem{Senz}
V. Senz, T. Heinzel, T. Ihn, S. Lindemann, K. Ensslin, W. Wegscheidez, 
and M. Bichler, Journal of Physics: Condensed Matter {\bf 13}, 
3831 (2001).

\bibitem{Cronenwett}
S. M. Cronenwett, H. J. Lynch, D. Goldhaber-Gordon, L. P. Kouwenhoven, 
C. M. Marcus, K. Hirose, N. S. Wingreen, and V. Umansky,
Phys. Rev. Lett. {\bf 88}, 226805 (2002).


\bibitem{Folk}
Y. Ren, W. W. Yu, S. M. Frolov, J. A. Folk, and W. Wegscheider, Phys. Rev. B {\bf 82}, 
045313 (2010).

\bibitem{Kogan}
A. Kogan, G. Granger, M. A. Kastner, D. Goldhaber-Gordon, and H. Shtrikman,
Phy. Rev. B {\bf 67}, 113309 (2003).

\bibitem{Roch}
Nicolas Roch, Serge Florens, Vincent Bouchiat, Wolfgang Wernsdorfer, 
and Franck Balestro, Nature {\bf 453}, 06930 (2008).

\bibitem{von}
J. Schmid, J. Weis, K. Eberl and K. v. Klitzing, Phys. Rev. Lett. {\bf 84},  5824 (2000). 

\end{references}
\end{document}